\documentclass[twocolumn,preprintnumbers,amsmath,amssymb]{revtex4}

\def\ga{\mathrel{\raise.3ex\hbox{$>$\kern-.75em\lower1ex\hbox{$\sim$}}}}
\def\la{\mathrel{\raise.3ex\hbox{$<$\kern-.75em\lower1ex\hbox{$\sim$}}}}

\newcommand\beq{\begin{equation}}
\newcommand\eeq{\end{equation}}
\newcommand\beqar{\begin{eqnarray}}
\newcommand\eeqar{\end{eqnarray}}
\usepackage{graphicx}
\usepackage{epsfig}
\usepackage{epsf}
\usepackage{rotating}

\begin{document}

\preprint{astro-ph/0309197}
\preprint{UMN--TH--2213/03}
\preprint{FTPI--MINN--03/24}
\vskip 0.2in
\title{The Chemical Evolution of Mg Isotopes vs. \\  the Time Variation of
the Fine Structure Constant}
\author{T. Ashenfelter}
\author {Grant J. Mathews}
\affiliation{Department of Physics, Center for Astrophysics, \\
University of Notre Dame, Notre Dame, IN, 46556}

\author{Keith A. Olive}
\affiliation{William I. Fine Theoretical Physics Institute, \\
University of Minnesota, Minneapolis, MN 55455, USA}

\begin{abstract}
The many-multiplet method applied to high redshift quasar absorption 
spectra has indicated a possible time variation of the fine structure constant.
Alternatively, a constant value of $\alpha$ is consistent with 
the observational analysis if a non-solar isotopic ratio of $^{24,25,26}$Mg
occurs at high redshift.  In particular, a higher abundance of the heavier isotopes
$^{25,26}$Mg are required to explain the observed multiplet splitting.  
We show that the synthesis of $^{25,26}$Mg  at the base of the convective
envelope in low-metallicity 
asymptotic giant branch stars, combined with a simple model of galactic
chemical evolution, can produce the required isotopic ratios and is supported by 
recent observations of high abundances of the neutron-rich Mg isotopes
in metal-poor globular-cluster stars. 
We conclude that the present data based on high redshift quasar absorption 
spectra may be providing interesting information on the nucleosynthetic 
history of such systems, rather than
a time variation of fundamental constants.
\end{abstract}

\maketitle



Over the last several years,  there has been considerable excitement
over the prospect that a time variation in the fine structure constant 
may have been observed in quasar absorption systems \cite{webb} - \cite{murphy3}.
While many
observations have led to interesting limits on the temporal variation
of $\alpha$ (see  \cite{uzan} for a recent review), only the many-multiplet method 
\cite{mm} has led to a positive result, 
namely that ${\delta \alpha \over \alpha}  = (0.54 \pm 0.12) \times 10^{-5}$
over a redshift range of $0.5 < z < 3$, where $\delta \alpha$ is defined as the present value
minus the past one.

 The sources of systematic errors in this method have been 
 well documented \cite{murphy2,murphy3} (see also \cite{bss}).
 Here, we would like to focus on one of these sources of systematic error
for which there is recent evidence of a new interpretation,  
namely the isotopic abundances of Mg assumed in the analysis.
 The analyses in \cite{webb} - \cite{murphy3}, have assumed
 terrestrial ratios for the three Mg isotopes.  They have also 
 shown that had they neglected the presence of the
neutron rich Mg isotopes, the case for a varying $\alpha$ would only be strengthened.
They further argued, based upon the galactic chemical evolution studies
available at that time, that the ratio of $^{25,26}$Mg/$^{24}$Mg is
expected to decrease at low metallicity making their result a robust and
conservative one.

In this paper,  we will show that it is in fact quite plausible that the
$^{25,26}$Mg/$^{24}$Mg ratio was sufficiently {\em higher} at low metallicity
to account for the apparent variation in $\alpha$.
As such, we would argue that while the many-multiplet method of analysis
does not unambiguously indicate a time variation in the fine structure constant,
it can lead to important new insights with regard to the nucleosynthetic history
of quasar absorption systems.

 
Regarding the observations of Mg isotopes, 
Gay and Lambert \cite{gl} determined the Mg isotopic ratios in 20 stars in the 
metallicity range $-1.8 <$ [Fe/H] $< 0.0$ with the aim of testing
theoretical predictions \cite{tww}. (The notation [A/B] refers to the the log
of the abundance ratio A/B {\em relative} to the solar ratio.) Their results confirmed 
that the $^{25,26}$Mg abundances relative to $^{24}$Mg 
appear to decrease at low metallicity for normal stars.
Although many stars were found to have abundance ratios somewhat higher than predicted,
even the `peculiar' stars which show enrichments in $^{25,26}$Mg
do not have abundance ratios substantially above solar.

Recently, however, a new study of Mg isotopic abundance in stars in the 
globular cluster NGC 6752 has been performed \cite{yong}. 
This study looked at 20 bright red giants which are all at a relatively 
low metallicity adopted at [Fe/H] = -1.62.  These observations show
a considerable spread in the Mg isotopic ratios which range from
$^{24}$Mg:$^{25}$Mg:$^{26}$Mg = 84:8:8 (slightly poor in the heavies)
to 53:9:39 (greatly enriched in $^{26}$Mg). The terrestrial value is
$^{24}$Mg:$^{25}$Mg:$^{26}$Mg = 79:10:11 \cite{rt}.  Of the 20 stars observed, 15
of them show $^{24}$Mg fractions of 78\% or less (that is below solar),
and 7 of them show fractions of 70\% or less with 4 of them in the range
53 - 67 \%. This latter range is low enough to have a substantial effect on a 
determination of $\alpha$ in quasar 
absorption systems if the same ratios were to be found there.
A previous study \cite{shetrone} also found unusually high abundances of
the heavy Mg isotopes in M13 globular-cluster giants.  Ratios of
$^{24}$Mg:$^{25,26}$Mg  were found as low as 50:50 and even 44:56.
Similar results were very recently found in \cite{yli}.
According to \cite{murphy3}, raising the heavy isotope concentration to
$^{24}$Mg:$^{25,26}$Mg = 63:37 would sufficiently shift the multiplet 
wavelengths to eliminate the need for a varying 
fine structure constant. 
While dispersion in the data could be a symptom of systematic errors
often occurring when data from several samples are combined,
real dispersion is a signal that the observed abundances 
have been affected by local events.

Available calculations of Type II supernova yields input into chemical 
 evolution models \cite{tww}, as well as observations of the Mg isotopic abundances
 in relatively low metallicity  stars, support the idea that the heavy Mg isotopes
 were rarer in the past.  Nevertheless, this conclusion is very sensitive to the
 the star formation history in object under consideration.  Intermediate mass
 stars in their giant phase, are expected to be efficient producers of $^{25,26}$Mg 
 \cite{fc,sll,lattanzio} and the recent data from a low-metallicity 
 globular cluster \cite{yong,yli} indicates  substantial amounts of variation in the 
 $^{25,26}$Mg/$^{24}$Mg ratios
 as well as a number stars highly polluted in the neutron rich Mg isotopes.
 These observations show isotopic ratios considerably higher than the
 ratios predicted in zero metallicity supernovae and they conclude 
 that asymptotic giant branch (AGB) stars may be responsible for this contamination.



Mg is produced in both Type I and Type II supernovae.
In Type II supernovae,  it is produced \cite{ww94} in the carbon and neon
burning shells with an abundance somewhat less than 10\% of 
the oxygen abundance produced in massive stars.  However, not much
$^{25,26}$Mg is produced in conventional stellar evolution at low metallicity.
This is because the isotopes $^{25,26}$Mg are produced primarily in the outer carbon layer
by the reactions $^{22}$Ne($\alpha$,n)$^{25}$Mg and
 $^{25}$Mg(n,$\gamma$)$^{26}$Mg.  According to the models
 of Woosley and Weaver \cite{ww94}, solar metallicity models
 produce final Mg isotope ratios reasonably close to 
 solar values, while more massive stars
 tend to be slightly enhanced in the heavy isotopes,
( e.g., the 25 M$_\odot$ model gives a ratio of 65:15:20).
 Furthermore, the abundance of $^{25,26}$Mg
 scales linearly with metallicity in the carbon shell. Hence, it would naively
 be expected that the ejecta from the first generation of supernovae would
 show a severe paucity of $^{25,26}$Mg.
 
 Much of the solar abundance of Mg is produced in Type Ia supernovae
 but with Mg to Fe ratios below solar.  For example, the models of Thielemann,
 Nomoto and Yokoi \cite{tny}, give [Mg/Fe] $\simeq -1.2$.  Due to the absence of free
 neutrons,  essentially no $^{25,26}$Mg are produced in Type Ia supernovae.

 The results of chemical evolution models tracing the Mg isotopes were
 presented in \cite{tww}.
 These models clearly show the effect of the low yields of $^{25,26}$Mg
 at low metallicity and predict  [$^{25,26}$Mg/$^{24}$Mg] $< -0.8$
 at [Fe/H] $< -1$. The result of \cite{tww} is essentially reproduced in the
dashed curve shown in Figure 1 and discussed below.

Based on the conventional theory of Mg production (e.g. \cite{ww94}, 
models of galactic chemical evolution \cite{tww}
and previously available data on the isotopic abundance of Mg \cite{gl}, the adoption
of solar isotopic Mg ratios by \cite{webb} - \cite{murphy3} in the many-multiplet analysis
would appear to be a safe and conservative one. 
These models indicate that the $^{24}$Mg:$^{25,26}$Mg
ratio was higher than 79:21 in the past and it was shown \cite{murphy3} that a higher 
ratio only strengthens the case for a varying $\alpha$. These models, however, do not include
contributions from intermediate mass stars as we now describe.
Recall that increasing
the abundances of the heavier Mg isotopes would yield a larger value
for $\alpha$, and a ratio of
$^{24}$Mg:$^{25,26}$Mg = 63:37 is sufficient to obviate 
the need for a varying 
fine structure constant.
 
Recently,  it has been appreciated \cite{fc,sll,lattanzio}
that intermediate mass stars of low metallicity  can also be efficient producers of
 the heavy Mg isotopes 
during the thermal-pulsing AGB  phase.  
Heavy magnesium isotopes (and to some extent silicon isotopes as well)
are synthesized via two mechanisms both of which are particularly
robust in 2.5-6 M$_\odot$ stars with low metallicity.
Such low-metallicity objects, indeed are 
precisely the kinds of objects which ought to produce the
abundances observed in QSO narrow-line absorption systems at high redshift.

One process is that of  hot-bottom burning.
This process is also 
believed to be a copious source
of lithium in low metallicity stars (cf. \cite{iwamoto})
During the AGB phase, stars develop an extended outer convective envelope.
Material in this  convective envelope is mixed downward to regions of high
temperature at the base.
Of particular interest for this paper is that 
the base of the envelope is more compact and of higher temperature 
in low metallicity stars than in
stars of solar composition.
This can be traced to the decreased opacity of these objects.
Furthermore, these stars would also have  a
shorter lifetime because they are hotter.  Low to intermediate mass stars
would contribute to the enrichment of the interstellar medium
 considerably sooner than their higher metallicity
conterparts.

Because these stars  become  sufficiently hot ($T \ge 7 \times 10^7$ K),
 proton capture processes in the Mg-Al cycle become effective. Proton capture
on $^{24}$Mg then leads to the production of $^{25}$Mg (from the decay of
$^{25}$Al)  and to $^{26}$Al (which decays
to $^{26}$Mg).

A second contributing process occurs deeper in the star
during  thermal pulses  of the helium-burning shell.
The helium shell experiences periodic thermonuclear runaways
when the ignition of the triple-alpha reaction occurs under electron-degenerate 
conditions.  Due to electron degeneracy, the star is unable to expand and cool.
Hence, the temperature rapidly rises until the onset of convection to transport 
the energy away.  During these thermal pulses,
$^{22}$Ne is produced by  $\alpha$
captures on $^{14}$N which itself is left over from the CNO cycle.
Heavy magnesium isotopes are then produced via  the   
$^{22}$Ne($\alpha$,n))$^{25}$Mg and   $^{22}$Ne($\alpha$,$\gamma$)$^{26}$Mg  reactions. 
It was argued recently \cite{sll}, that in intermediate mass stars which
experience a 3rd dredge-up, significantly greater amounts of $^{25,26}$Mg are produced.  
A key point is that even though seed material is less plentiful in low metallicity
stars, the reactions are very temperature sensitive.  Hence, the
increased temperature in the interior of low-metallicity stars more than
compensates for the depleted seed material, leading to significant production
of the heavy Mg isotopes. It has even been argued that these processes
may also be net destroyers of $^{24}$Mg \cite{fc,lattanzio} due to the extreme temperatures
attained.

To illustrate the effects of producing enhanced abundances of $^{25,26}$Mg
in intermediate mass stars, we show the results of a simple model of 
galactic chemical evolution which traces the Mg isotopic abundances
with and without the AGB source.  When combined with
the recent data showing such enhancements, our results
suggest a plausible alternative for the interpretation of the 
quasar absorption system data based on the many-multiplet method.


For our purposes a
simple recalculation of the results of Timmes et al. \cite{tww} with
and without the contribution from intermediate-mass AGB stars is sufficient.
This allows us to make a direct comparison with the conclusions 
of the previous authors.

The galactic chemical evolution model of Timmes et al.  \cite{tww} is based upon  exponential 
infall and a Schmidt star formation rate.
We utilize a slightly modified 
model with updated yields \cite{up1} which nevertheless reproduces the results of \cite{tww}
in the appropriate limit.
Hence, we write 
the evolution of the surface density $\sigma_i$ of an  isotope $i$
as,
\begin{eqnarray}
{d \sigma_i \over dt} =  
\int_{0.8}^{40} B(t - \tau(m)) \Psi(m) X_i^S(t-\tau(m)) dm
\nonumber \\
+ \int_{2.5}^{9.0} B(t - \tau(m)) \Psi(m) X_i^{AGB}(t-\tau(m)) dm
\nonumber \\
+ m_{CO} X_i^{Ia}R_{Ia} -B(t) {\sigma_i \over \sigma_{gas}} + \dot \sigma_{i,gas},
\label{sigdot}
\end{eqnarray}
where $B(t)$ is the stellar birthrate at $t$,
$\Psi(m)$ is the initial mass function (IMF), $X_i^S$ is the mass fraction of isotope $i$ 
ejected from single star evolution and Type II supernovae, and
$\tau(m)$ is the lifetime of a star of mass $m$. 
The 2nd term in Eq. \ref{sigdot}
is the new contribution from AGB stars.  For this purpose we adopt the AGB 
yields of \cite{lattanzio}.  The 3rd term in Eq. \ref{sigdot} is the contribution from
type Ia supernovae where $m_{CO}$ is the mass the exploding carbon-oxygen 
white dwarf, and $R_{Ia}$ is the SNIa supernova rate taken from \cite{kobayashi}. 
Finally, the last two terms represent the trapping 
of elements in new stars, and the galactic infall rate (presumed to be of primordial material). 

As noted by a number of authors, the yields of heavy magnesium isotopes in AGB stars
is extremely temperature sensitive, and hence rather sensitive to
detailed physics of the stellar models.   Moreover, there are reasons to expect that the initial
mass function at low metallicity could be biased toward intermediate-mass stars.
One argument for this is simply that with fewer metals, the cooling is less efficient
in the protostellar cloud, so that a more massive cloud is required to form a star.
To account for these possibilities, we introduce a modest enhancement of the IMF
for intermediate mass stars.  Such an enhancement has often been proposed
and is motivated by models for star formation at low metallicity. For example, it
has been invoked \cite{wd} to account for a large population of white dwarfs as 
microlensing objects in the Galactic Halo.
In fact, such a population of intermediate mass stars was recently proposed \cite{foscv} to 
explain the dispersion of D/H observed in quasar absorption systems \cite{kirk}.
Hence for the creation function, $B\Psi$, we write:
\begin{eqnarray}
\label{imf}
B(t)\Psi(m) & = &B_1(t) \Psi_1(m) + B_2(t) \Psi_2(m) \qquad     \\
= B_1 m^{-2.35} &  + &(B_2/m) \exp{(-\log{(m/5.0)}^2)/(2 \sigma^2)}). \nonumber
\end{eqnarray}
The IMF in Eq.~(\ref{imf}) accounts for a standard Salpeter distribution of stellar masses,
$\Psi_1(m)$, with the addition of a lognormal component of 
stars peaked at 5 M$_\odot$, $\Psi_2$. The dimensionless width, $\sigma$, is taken to be 0.07.
This IMF is similar to a Gaussian with a width of $5\sigma = 0.35$ M$_\odot$.
For the normal stellar component we take the time dependence as
\begin{equation}
B_1(t) = (1.0-e^{-t/.5{\rm Gyr}})\sigma_{tot}(t) [{\sigma_{gas} / \sigma_{tot}(t)}]^2~~,
\end{equation}
while for the intermediate mass component we take
\begin{equation}
B_2(t) = 5.5 e^{-t/.2{\rm Gyr}}\sigma_{tot}(t) [{\sigma_{gas} / \sigma_{tot}(t)}]^2~~.
\end{equation}
This is very similar to the model used in \cite{foscv}.  It contains an early burst
of intermediate mass stars peaked at 5 M$_\odot$ which is exponentially
suppressed after 0.2 Gyr.  The second component describes standard quiescent
star formation with a smooth transition from the burst.

Figure 1 shows a comparison of our calculated magnesium isotope ratio vs iron abundance.
The solid curve shows the result of the model described above including the AGB contribution.
The QSO absorption-line systems in question have
metallicities in the range from 0 to -2.5 with a typical iron abundance of
 [Fe/H]$ \sim -1.5$. 
The mean isotopic ratio needed to account for the data of \cite{webb}-\cite{murphy3} is
 $^{25,26}$Mg/$^{24}$Mg = 0.58 (shown by the solid horizontal line)
 with a $1\sigma$ lower limit of 0.47 (dashed horizontal line).
  This figure clearly demonstrates that a plausible model
is possible in which a sufficient abundance of heavy Mg isotopes can be produced 
to both explain the
observed globular-cluster data and the apparent trends in the many-multiplet data
or QSO absorption-line systems  at
high redshift. 

The behavior in the evolution of the heavy isotopes can be explained as follows:
Initially, the production of  $^{25,26}$Mg in the ejecta of intermediate mass
stars is delayed by their relatively long lifetime (compared to very massive stars).
Initial contributions to the chemical enrichment of the interstellar medium 
comes from the most massive and shortest lived stars.  In this model, the burst of 
intermediate mass stars begins to produce  $^{25,26}$Mg at [Fe/H] $\ga -2.5$.
At this stage, during the intermediate mass burst, $^{25}$Mg and $^{26}$Mg are
copiously produced relative to $^{24}$Mg as per the yields of \cite{lattanzio}. 
At higher metallicity, the ejecta from the standard population of (massive) stars
which is poor in $^{25,26}$Mg begins to dilute the ratio relative to $^{24}$Mg,
thereby producing the noticable bump in $^{25,26}$Mg/$^{24}$Mg around
[Fe/H] $\sim -1.5$. At late times, the effect of the early generation of intermediate
mass stars is largely washed away. 

The dashed curve excludes the AGB yields and the intermediate mass component. 
It gives a result similar to that of \cite{tww}.
We note that recently the AGB contribution was included in a chemical evolution model
\cite{fenner} for a normal stellar distribution.  While the results showed significantly higher
abundances of $^{25,26}$Mg relative to $^{24}$Mg than that given by the dashed curve, 
they were not high enough to account for the claimed variability in $\alpha$.  
An enhanced early  population of intermediate mass stars is therefore necessary.
Nevertheless, this seems more plausible than a time varying fundamental
constant.

\begin{figure}
\mbox{\epsfig{file=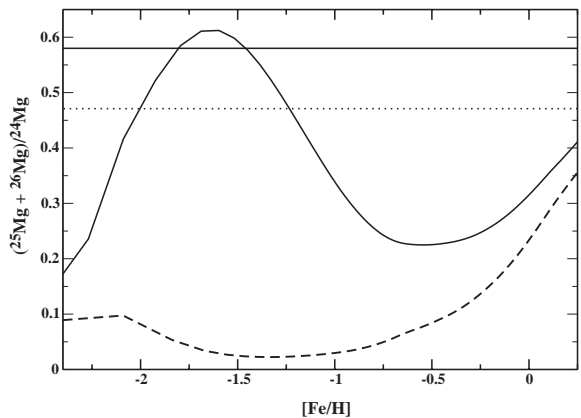,height=5.5cm,angle=0}}
\caption{\label{fig:epsart} The chemical evolution of the  $^{25,26}$Mg
isotopes relative to  $^{24}$Mg. The solid curve is our result based on 
eq. (\protect\ref{imf}) using the AGB Mg yields of \protect\cite{lattanzio}.  The dashed curve
is the result of turning off the AGB contribution and excluding the burst of intermediate
mass stars.  The horizontal lines indicate the ratio of $^{25,26}$Mg/$^{24}$Mg
necessary to explain the shifts seen in the many-multiplet analysis.}
\end{figure}


We have argued that previous models for the apparent broadening of the
Mg multiplet in QSO absorption-line systems may have left out the
important possible contribution from the production of heavy magnesium
isotopes during the AGB phase of low-metallicity intermediate-mass stars.
We have shown that a simple, plausible galactic chemical evolution model 
can be constructed which explains both the large abundances of heavy Mg isotopes
observed in globular clusters and the large abundance necessary to
explain the many-multiplet data.  We note that the hot bottom burning process in
AGB stars is also likely to  have altered the  Si isotopes as well by proton captures
on aluminum and silicon at the base of the convective envelope.
Hence, it is possible that the supporting data from Si isotopes can be explained by this
paradigm as well.  Obviously more detailed work is warranted to clarify the
ability of this mechanism to account for the data.  Nevertheless,
the model presented here is based upon plausible expectations of stellar and galactic
evolution and should be taken seriously before demanding an alteration of any fundamental
constant at high redshift.  


We thank C. Cardall for helpful conversations.
The work of K.A.O. was partially supported by DOE grant
DE--FG02--94ER--40823.
 Work at the University of Notre Dame was supported by the
U.S.~Department of Energy
under Nuclear Theory Grant DE-FG02-95-ER40934.

\end{document}